\documentclass[aps,prc,twocolumn,superscriptaddress,nofootinbib,tightenlines,
preprintnumbers,showkeys,floatfix]{revtex4-2}

\usepackage[utf8]{inputenc}
\usepackage[english]{babel}
\usepackage{amssymb,amsthm,amsmath,amstext,amsbsy,amsopn,mathrsfs}
\usepackage{bbm}
\usepackage{nicefrac}
\usepackage{slashed}
\usepackage{graphicx}
\usepackage{hyperref}
\usepackage{leftidx}
\usepackage{environ}
\usepackage{mathtools}
\usepackage{xspace}
\usepackage{array}
\usepackage{xcolor}
\usepackage{pgfplotstable}
\usepackage{isotope}
\usepackage{suffix}
\usepackage{booktabs,colortbl}
\usepackage{physics}
\usepackage{braket}

\newcommand{\ie}{\textit{i.e.}\xspace}

\newcommand{\apriori}{\textit{a priori}\xspace}
\WithSuffix\newcommand\apriori*{\textit{a-priori}\xspace}

\newcommand{\viz}{\textit{viz.}\xspace}

\newcommand{\mathspace}{\ \ }
\newcommand{\mathtext}[1]{\mathspace\text{#1}\mathspace}

\renewcommand{\vec}[1]{\mathbf{#1}}

\newcommand{\vecr}{\mathbf{r}}
\newcommand{\vecx}{\mathbf{x}}

\newcommand{\vecn}{\mathbf{n}}

\newcommand{\vecp}{\mathbf{p}}

\newcommand{\ii}{\mathrm{i}}
\newcommand{\ee}{\mathrm{e}}

\newcommand{\OO}{\mathcal{O}}
\newcommand{\ZZ}{\mathbb{Z}}

\newcommand{\CC}{\mathbb{C}}

\newcommand{\Rp}{\mathrm{Re}}
\newcommand{\Ip}{\mathrm{Im}}

\newcommand{\psiinf}{\psi_\infty}

\newcommand*\rvec[1]%
{\ensuremath{\overset{\smash{\raisebox{-1.5pt}{\tiny$\rightarrow$}}}{#1}}}
\newcommand*\lvec[1]%
{\ensuremath{\overset{\smash{\raisebox{-1.5pt}{\tiny$\leftarrow$}}}{#1}}}

\NewEnviron{subalign}[1][]{%
\begin{subequations}\begin{align}
  \BODY
\end{align}\label{#1}\end{subequations}
}

\NewEnviron{spliteq}{%
\begin{equation}\begin{split}
  \BODY
\end{split}\end{equation}
}

\NewEnviron{mleq}{%
\begin{equation}
  \BODY
\end{equation}
}

\newcommand{\dvrsum}[2]{\sum\limits_{#1={-}#2/2}^{#2/2-1}}

\newcommand{\cscale}{\zeta}
\newcommand{\cc}[1]{{#1}^*}

\newcommand{\cabs}[1]{\abs{#1}_{\text{cs}}}

\begin{document}

\title{Complex scaling in finite volume}

\author{Hang Yu}
\email{yhang@ncsu.edu}
\affiliation{Department of Physics, North Carolina State University,
Raleigh, North Carolina 27695, USA}

\author{Nuwan Yapa}
\email{ysyapa@ncsu.edu}
\affiliation{Department of Physics, North Carolina State University,
Raleigh, North Carolina 27695, USA}

\author{Sebastian König}
\email{skoenig@ncsu.edu}
\affiliation{Department of Physics, North Carolina State University,
Raleigh, North Carolina 27695, USA}

\begin{abstract}
Quantum resonances, \ie, metastable states with a finite lifetime, play an
important role in nuclear physics and other domains.
Describing this phenomenon theoretically is generally a challenging task.
In this work, we combine two established techniques to address this challenge.
Complex scaling makes it possible to calculate resonances with
bound-state-like methods.
Finite-volume simulations exploit the fact that the infinite-volume properties
of quantum systems are encoded in how discrete energy levels change as one
varies the size of the volume.
We apply complex scaling to systems in finite periodic boxes and derive the
volume dependence of states in this scenario, demonstrating with explicit
examples how one can use these relations to infer infinite-volume resonance
energies and lifetimes.
\end{abstract}

\maketitle

\section{Introduction}
\label{sec:Introduction}

A collection of results going back to the groundbreaking early work of
Lüscher~\cite{Luscher:1985dn,Luscher:1986pf,Luscher:1990ux} makes it possible to
infer the properties of quantum systems from simulations in finite periodic
boxes.
The essence of the technique is that real-world properties of a system are
encoded in how its discrete energy levels change as volume size is varied.
For example, bound-state energy levels depend exponentially on the volume,
with a scale that is set by the momentum corresponding to the energy
relative to the nearest two-cluster breakup~\cite{Konig:2017krd}, and the
prefactor in this dependence is proportional to the asymptotic
normalization coefficient (ANC) corresponding to that channel.
Information about elastic scattering, on the other hand, can be determined from
energy levels with power-law behavior, via the Lüscher quantization
condition~\cite{Luscher:1990ux}.
Extending the repertoire of such relations is a field of very active research,
with focus in recent years in particular on three-body
systems~\cite{Kreuzer:2010ti,Kreuzer:2012sr,Polejaeva:2012ut,Briceno:2012rv,%
Kreuzer:2013oya,Meissner:2014dea,Hansen:2015zga,Hammer:2017uqm,%
Hammer:2017kms,Mai:2017bge,Doring:2018xxx,Pang:2019dfe,Culver:2019vvu,%
Briceno:2019muc,Romero-Lopez:2019qrt,Hansen:2020zhy,Muller:2021uur,%
Draper:2023xvu,Bubna:2023oxo}.
Lattice quantum chromodynamics (lattice QCD) is the primary domain where these
methods are currently applied, but they can also be used in connection
with lattice effective field theory (lattice EFT)
calculations of atomic nuclei~\cite{%
Lee:2008fa,Lahde:2019npb,Lu:2021tab,Shen:2022bak,Elhatisari:2022qfr}
and other few-body approaches~\cite{Dietz:2021haj,Bazak:2022mjh}.

Resonances, \ie., quasibound states that decay with a finite lifetime, are
manifest in the finite-volume spectrum as avoided crossings between states, and
for two-body systems it is straightforward to associate this feature with a
steep rise in the scattering phase
shift~\cite{Wiese:1988qy,Luscher:1991cf,Rummukainen:1995vs}.
It has also been shown that this feature carries over to systems of more than
two particles that host few-body resonances~\cite{Klos:2018sen}.
While this makes finite-volume calculations an interesting tool to detect the
presence of few-body resonance states (or corroborate their absence), this
method does not provide a straightforward way to quantitatively determine the
``width'' (proportional to the inverse lifetime) of few-body resonances.
Moreover, since the avoided crossings are expected to appear among
``scattering'' levels with power-law volume dependence, identifying low-energy
resonances generally requires calculations in large boxes, which can be
numerically very expensive.
Finite-volume eigenvector continuation has been developed to reduce
this cost ~\cite{Yapa:2022nnv}, but it is still very interesting and relevant to
look for alternative methods that are able to determine resonance properties
comprehensively (\ie, which also give access to the decay width), and which
are ideally at the same time more efficient in terms of numerical cost.
In this work, we develop such an alternative by combining the finite-volume
approach with the so-called ``complex scaling'' method.

Although resonances are inherently a time-dependent phenomenon (they decay after
existing for a finite time), techniques exist that enable their description
within the framework of time-independent scattering theory.
A key quantity for this theory is the so-called scattering matrix ($S$
matrix), which can be considered as a function of a \emph{complex} energy
$E$ defined on multiple Riemann sheets.
Standard phenomena like scattering and bound states appear for real $E$ on the
first (``physical'') sheet, while (decaying) resonances are manifest as poles
of the $S$ matrix at complex energies $E = E_R - \ii\Gamma/2$ on the second
Riemann sheet with $E_R$ the resonance position and $\Gamma$ the
width.\footnote{In this brief description we tacitly assume that we are
discussing a two-body system.
For more particles and/or multichannel problems, the Riemann-sheet structure
becomes richer.}
If $\Gamma$ is not large compared to $E_R$, these poles appear close to the
scattering regime and therefore lead to the characteristic peaks in the cross
section that resonances are commonly associated with phenomenologically.
Within this quasistationary formalism, the $S$-matrix resonance poles are
associated with complex-energy
eigenstates~\cite{Baz:1969,Gamow:1928zz,Siegert:1939zz}.

Accessing these poles (or equivalently the corresponding complex energy
eigenstates) is generally nontrivial.
Clearly, a non-Hermitian extension of the formalism is necessary to
accommodate such states because for systems described by a Hermitian
Hamiltonian, all eigenstates must have real energy eigenvalues.
One way of achieving this extension is the so-called ``complex scaling method
(CSM)''~\cite{Reinhardt:1982aa,Moiseyev:1998aa},
described further below, which in this work we formulate in a periodic finite
volume (FV).
We note that this is closely related to the approach of Ref.~\cite{Guo:2020ikh},
which has shown how resonance properties can be obtained from transition
amplitudes calculated in finite volume after analytic continuation to purely
imaginary box sizes.
By instead applying complex scaling to the Hamiltonian of a
system, we are able to obtain resonance energies directly as complex energy
eigenvalues via diagonalization.
Moreover, we study how these resonance energies depend on the size of the
volume, which is similar to bound states, but inherently richer because
both the real part and the imaginary part of the energy exhibit volume
dependence.
We derive in detail the functional form of this volume dependence for
two-cluster states.
Moreover, we describe a concrete numerical implementation for calculating
generic complex-scaled few-body systems in finite volume and use this to
demonstrate with explicit examples how our analytical relations can be used to
determine infinite-volume resonance positions (the real part of the resonance
energy) and the associated widths (given by the imaginary part) from a range of
finite-volume simulations.

Our paper is organized as follows.
In the following Sec.~\ref{sec:Formalism}, we first introduce
the CSM in general and then proceed to
discuss the volume dependence arising from imposing periodic boundary conditions
on complex-scaled resonance states.
In Sec.~\ref{sec:Numerical} we describe our numerical implementation
and we proceed to study a series of explicit examples in Sec.~\ref{sec:Examples}.
We close in Sec.~\ref{sec:Conclusion} with a summary and and outlook.

\section{Formalism}
\label{sec:Formalism}

\subsection{Complex scaling method}
\label{sec:ComplexScaling}

The (uniform) complex-scaling method~\cite{Aguilar:1971ve,Balslev:1971vb,
Moiseyev:1978aa,Moiseyev:1998aa,Afnan:1991kb,ho_method_1983,Reinhardt:1982aa}
makes it possible to describe resonances in a way that is very similar to
bound-state calculations.
This is achieved by expressing the wave function not along the usual real
coordinate axis, but instead along a contour rotated into the complex plane.
For example, if $r$ denotes the relative distance between particles in a
two-body system, described by a Hamiltonian
\begin{equation}
 H = H_0 + V
 \label{eq:H-generic}
\end{equation}
with free (kinetic) part $H_0$ and interaction $V$, then complex scaling is
implemented by applying the transformation
\begin{equation}
 r \to r \ee^{\ii \phi} \equiv r\cscale
 \label{eq:r-scaled}
\end{equation}
with some angle $\phi$, the appropriate choice of which in general depends on
the position of the resonance one wishes to study: if the state of interest has
a complex energy $E$, then it is necessary to ensure that $\phi >
{-}\frac{\arg{E}}{2}$.
If one were to solve the Schrödinger equation in differential form without
complex scaling while imposing boundary conditions that are appropriate for
resonance states, the resulting wave function would have an amplitude that
grows exponentially with $r$ and does therefore not describe a normalizable
state.
More specifically, for a two-body state with energy $E$ corresponding to an
$S$-matrix pole, the asymptotic behavior of the radial wave function for large
separation $r$ between the two particles is given
by~\cite{Taylor:1972,Faldt:1996sd}
\begin{equation}
 \psi(r)
 \xrightarrow[r \rightarrow \infty]{} N \, \hat{h}^+_l(k r)
 \sim \exp({-}\kappa r) \,,
\label{eq:psi-asymptotic}
\end{equation}
where $k = \sqrt{2\mu E}$ is the associated momentum scale (with $\mu$ the
reduced mass of the system), and $l$ denotes the angular momentum of the state.
The function $\hat{h}^+_l(kr)$ is a Riccati-Hankel function, the dominant
behavior of which for large argument is exponential (times an $l$-dependent
polynomial that we omitted in Eq.~\eqref{eq:psi-asymptotic} for simplicity).

Clearly, when $E$ lies in the fourth (lower right) quadrant of the complex
plane, so does the corresponding momentum $k$, and then the imaginary part
causes the Riccati-Hankel function to \emph{grow} exponentially.
What is accomplished with the transformation~\eqref{eq:r-scaled} is that along
the rotated contour, the same (now analytically continued) wave function behaves
similar to a bound state, \ie, its amplitude exponentially tends to zero as
$r\to\infty$.

Complex scaling can be further elucidated by considering the same system in
momentum space.
The scaling of the radial coordinate $r$ is equivalent to a rotation in momentum
representation that goes in the opposite (clockwise) direction with the same
angle $\phi$~\cite{Afnan:1991kb}, \ie, if we consider the wave function in terms
of a momentum coordinate $q$ conjugate to $r$, then complex scaling is
implemented as
\begin{equation}
 q \to q \ee^{{-}\ii \phi} = q\cc{\cscale} \,.
\end{equation}
Alternatively, this scaling in momentum space can be understood as a rotation of
the $S$-matrix branch cut in the complex-energy plane by an angle $2\phi$
clockwise, thereby exposing a section of the second Riemann sheet where
resonances are located~\cite{Afnan:1991kb}.
Choosing $\phi$ sufficiently large, as mentioned above, then corresponds to
``revealing'' enough of the second sheet to uncover the resonance pole.

In the remainder of this subsection, we address several technical aspects
that are relevant for the concrete finite-volume implementation of the complex
scaling method that we consider in this work.

\paragraph*{Three-dimensional Cartesian coordinates.}
While the method of complex scaling is easier to explain in a partial-wave
framework~\cite{Afnan:1991kb,ho_method_1983},
the equivalent three-dimensional (3D) Cartesian formulation, which is most appropriate for the
cubic-box geometry we study in this paper, needs to be carefully stated.
To that end we note that complex scaling of each individual component
of $\vec{r}=(x,y,z)$ is equivalent to complex scaling of the radial
coordinate $r$,
\begin{multline}
 r = \sqrt{x^2+y^2+z^2} \\
 \to \sqrt{(x\cscale)^2+(y\cscale)^2+(z\cscale)^2} = r\cscale \,,
\label{eq:radial_coordinate_scaling}
\end{multline}
but it leaves the angles $\theta$ and $\varphi$ in spherical coordinates
$\vecr = (r,\theta,\phi)$ unaffected:
\begin{subalign}
 \cos{\theta} &= \frac{z}{r} = \frac{z\cscale}{r\cscale} \,, \\
 \tan{\varphi} &= \frac{y}{x} = \frac{y\cscale}{x\cscale} \,.
\end{subalign}
Therefore, we can apply complex scaling directly to Cartesian coordinates
as long as Eq.~\eqref{eq:radial_coordinate_scaling} is used to calculate
the corresponding radial distances.
For later use, we define a version of the Euclidean norm that preserves
complex scaling as in Eq.~\eqref{eq:radial_coordinate_scaling}, \viz
\begin{equation}
 \cabs{\vec{r}} = \sqrt{x^2+y^2+z^2}
 \mathtext{for} \vec{r}=(x,y,z), \;
 x,y,z \in \CC \,.
\label{eq:cabs}
\end{equation}

\paragraph*{Relative coordinates.}
In line with the numerical implementation for few-body systems in a box
that we describe further in Sec.~\ref{sec:Numerical}, we consider now a
system of $n$ particles described in terms of simple relative coordinates,
which we define as
\begin{equation}
 \vec{x}_i = \begin{cases}
  \vec{r}_i - \vec{r}_n & \text{for}\ 1 \leq i < n \,, \\
  \frac{1}{n} \sum_{j=1}^n \vec{r}_j & \text{for}\  i=n
 \end{cases}
\label{eq:simple_relative_coordinates}
\end{equation}
in terms of the single-particle coordinates $\vec{r}_i$, $i=1,\ldots ,n$.
Note that $\vec{x}_n$ in this notation is the overall center-of-mass coordinate
that does not appear explicitly in the description of translationally invariant
systems.
Complex scaling can be applied simultaneously to each of these
relative coordinates.
That is, for each $\vec{x}_i$ we can simply apply the transformation
\begin{equation}
\vec{x}_i \to \Vec{x}_i \ee^{\ii \phi} \equiv \vec{x}_i \cscale \,,
\end{equation}
and from the previous discussion we know that this is equivalent to scaling
each radial modulus $x_i \equiv \abs{\vec{x}_i}$ as $x_i \to \cscale x_i$.
If we consider for simplicity a system with only local pairwise two-body
interactions that are spherically symmetric, then each potential term in the
Hamiltonian is transformed as
\begin{equation}
 V(\vec{x}_i) = V(x_i) \to V(\cscale x_i) \,,
\end{equation}
or, for interacting pairs that are not directly described by one of the
$\vec{x}_i$, as
\begin{equation}
 V(\abs{\vec{x}_i - \vec{x}_j})
 \to V(\cscale \abs{\vec{x}_i - \vec{x}_j})
 \mathtext{,} i \neq j \,,
\label{eq:V-xixj}
\end{equation}
which follows from the fact that we scale each $\vec{x}_i$ with the same
rotation angle.
Alternatively, we can state the prescription that in order to evaluate
interactions, relative distances should be evaluated using the Euclidean
``norm'' as defined in Eq.~\eqref{eq:cabs}, preserving the rotation
angle for complex coordinates, \ie,
$V(\abs{\vec{x}_i - \vec{x}_j}) \to V(\cabs{\vec{x}_i - \vec{x}_j})$.
Either way, complex scaling for the interaction implies that we consider
the analytic continuation of $V$ from real coordinates to complex-scaled ones.

The kinetic-energy operator (free Hamiltonian) can be expressed in terms
of second derivatives with respect to the Cartesian components of
the $\vec{x}_i$:
\begin{equation}
 K = {-}\frac{1}{2\mu} \sum_{i=1}^{n-1} \sum_{j=1}^{i} \sum_{c=x,y,z}
 \partial_{i,c} \partial_{j,c}
 \mathtext{m}
 \partial_{i,c} = \frac{\partial}{\partial x_i^{(c)}} \,.
\label{eq:K}
\end{equation}
This includes some mixed-derivative terms because we are using simple relative
coordinates, but since the complex-scaling phase is the same for each
$\vec{x}_i$, it is clear that ultimately we have
\begin{equation}
 H_0 \to \ee^{-2\ii\phi} H_0 = (\cscale^*)^2 H_0
\label{eq:H-0-scaled}
\end{equation}
under complex scaling.
Note that this behavior would be the same for any other relative coordinate
system, such as Jacobi coordinates, as long as the complex scaling can be
expressed as arising from a scaling of each single-particle coordinate
$\vec{r}_i$, $i=1,\ldots ,n$, with a uniform angle $\phi$.

\subsection{Volume dependence}
\label{sec:VolumeDependence}

We now consider a two-body state generated by
a Hamiltonian $H = H_0 + V$ with kinetic part $H_0$ and a short-range
interaction $V$ that becomes negligible when the particles are separated by more
than a distance $R$.
For a bound-state with energy $E_\infty = {-}\kappa_{\infty}^2/(2\mu)$
in infinite volume, considered within a cubic geometry with periodic
boundary conditions (periodic box), the binding energy becomes a function
of the edge length $L$ of the box.
The leading form of this volume dependence is known to be given by
\begin{multline}
 \Delta E(L) \equiv E(L) - E_\infty \\
 = \frac{3 \gamma_\infty^2}{\mu}
 \frac{\exp({-}\kappa_\infty L)}{L}
 + \OO\left(\ee^{{-}\sqrt{2}\kappa\cscale L}\right) \,,
\label{eq:LOform-BS}
\end{multline}
where $\gamma_\infty$ denotes the ANC of the bound state.
As discussed for example in
Refs.~\cite{Luscher:1985dn,Konig:2011nz,Konig:2011ti},
Eq.~\eqref{eq:LOform-BS} can be derived by making an ansatz
\begin{equation}
 \psi_{L,0}(\vecx)
 = \sum_{\vecn\in\ZZ^3} \psi_\infty(\vecx + \vec n L)
\end{equation}
for the wave function of the state at volume $L$, where $\psi_\infty(\vecx)$
denotes the states wave function in infinite volume.
In order to address some subtle points associated with complex scaling,
in the following we work through the analog of this approach for a
complex-scaled one-dimensional (1D) system system.
Following this, we comment briefly on the extension of the 1D method to the
3D system, which we then proceed to discuss in
detail using a more abstract method that has the advantage of giving
access to important subleading corrections.

\subsubsection{Leading volume dependence}

Let $\psi_\infty(\cscale x)$ be the complex-scaled wave function of a resonance
state in infinite volume, with energy $E_\infty = E(\infty)$ and associated momentum
$p_\infty$.
We can closely follow the derivation of the volume dependence for bound
states and start from the following ansatz for the state's wave function
when subject to an $L$-periodic boundary condition:
\begin{equation}
 \psi_{\cscale L,0}(x)
 = \sum_{n={-}\infty}^\infty \psi_\infty(\cscale x + \cscale nL) \,.
\end{equation}
Note that we use a subscript $\cscale L$ here to indicate explicitly
that this is the complex-scaled finite-volume ansatz, and for convenience we
define $\psi_{\cscale L,0}(x)$ so that its argument is explicitly
\emph{real} again.
Importantly, the shifts in the wave function are applied along the
rotated contour.
By construction, $\psi_{\cscale L,0}(x)$ satisfies
\begin{equation}
 \psi_{\cscale L,0}(x + nL) = \psi_{\cscale L,0}(x)
\end{equation}
for any $n\in\ZZ$, and the same must be true for the exact wave function at
volume $L$, which we denote as $\psi_{\cscale L}(x)$, also with real argument
defined along the rotated axis.

At this point we also assume for convenience that the interaction $V$ is a
simple local potential and note that general non-local potentials can be
considered analogously to the derivation in Ref.~\cite{Konig:2011ti}.
The complex-scaled finite-volume Hamiltonian $H_{\cscale L}$ is then obtained by
making the potential periodic, \viz
\begin{equation}
 V(\cscale x) \to V_{\cscale L}(x)
 \equiv \sum_{n={-}\infty}^\infty V(\cscale x + \cscale n L) \,,
\end{equation}
along with scaling the kinetic part as in Eq.~\eqref{eq:H-0-scaled}.
Acting with $H_{\cscale L}$ on $\psi_{\cscale L,0}(x)$, we find that
\begin{multline}
 H_{\cscale L} \psi_{\cscale L,0}(x) = E(\infty) \psi_{\cscale L,0}(x) \\
 \null + \sum_n\sum_{n'\neq n } V(\cscale x + \cscale nL)
 \psi_\infty(\cscale x+ z nL) \\
 \equiv E(\infty) \psi_{\cscale L,0}(x) + \eta(x) \,.
\end{multline}
Since the amplitude of the complex-scaled resonance wave function decays
exponentially like a bound state, we find that the function $\eta$ defined
above behaves as $\eta(\cscale x) \sim \OO (\ee^{\ii \cscale p_\infty L})$.
We can choose $\beta$ so that $\beta \ket{\psi_{\zeta L,0}}$ differs from the true
finite-volume wave function $\ket{\psi_{\cscale L}}$ only by an orthogonal term,
\ie, for
\begin{equation}
 \ket{\psi'_{\cscale L}}
 \equiv \ket{\psi_{\cscale L}}-\beta\ket{\psi_{{\cscale L},0}}
\end{equation}
it holds that $\braket{\psi_{{\cscale L},0}|\psi'_{\cscale L}} = 0$.
We have switched here to bra-ket notation for convenience and note that in
evaluating overlaps and matrix elements, the so-called
``c-product''~\cite{Moiseyev:1978aa,Moiseyev:2011} needs to be used, \ie,
the wave functions arising from bra states are not complex-conjugate when
evaluating inner products.

We now consider the matrix element
$\braket{\psi_{\cscale L}|H_{{\cscale L}}|\psi_{{\cscale L},0}}$
and let the Hamiltonian act to both left and right,
which gives
\begin{multline}
 \beta E(L)\braket{\psi_{\cscale L}|\psi_{{\cscale L},0}} \\
 = \beta E(\infty) \braket{\psi_{\cscale L}|\psi_{{\cscale L},0}}
 + \braket{\psi_{\cscale L}|\eta} \,,
\end{multline}
where $E(L)$ is the energy at volume $L$.
Noting that $\braket{\psi_\phi|\psi_{z,0}}
= \braket{\psi_{\phi,0}|\psi_{\phi,0}}$, we then find the finite-volume
energy shift as
\begin{multline}
 \Delta E(L) = E(L) - E_\infty \\
 = \frac{\braket{\psi_{\cscale L}|\eta}}
   {\beta\braket{\psi_{{\cscale L},0}|\psi_{{\cscale L},0}}}
 = \frac{\braket{\psi_{{\cscale L},0}|\eta}}
   {\braket{\psi_{{\cscale L},0}|\psi_{{\cscale L},0}}}
+ \frac{\braket{\psi_{\cscale L}'|\eta}}
   {\beta\braket{\psi_{{\cscale L},0}|\psi_{{\cscale L},0}}} \,.
\label{eq:shift-groundup}
\end{multline}
It can be shown that
$\braket{\psi_{\cscale L}'|\eta} = \OO(\ee^{\frac32 \ii \cscale p_\infty L})$
and is exponentially suppressed, using the asymptotic behavior of the
complex-scaled wave function in infinite volume.
We keep in mind here that $p_\infty = \sqrt{2\mu E_\infty}$, and with $E_\infty$
lying in the fourth quadrant of the complex energy plane, so does $p_\infty$.
Multiplication with $\cscale = \exp(\ii\phi)$ where $\phi> \arg p_\infty$
then ensures that $\ii\cscale p_\infty L$ has a negative real part, and
therefore indeed $\braket{\psi_\phi'|\eta}$ does not contribute to the
leading term.
Furthermore, on the domain $x\in [{-}L/2,L/2]$, we have that
\begin{equation}
 \eta(x) = V(\cscale x) \psi_\infty(\cscale x - \cscale L)
 + \OO (\ee^{\frac32 \ii \cscale p_\infty L}) \,.
\end{equation}
Putting this back into Eq.~\eqref{eq:shift-groundup}, we get
\begin{multline}
 \Delta E(L) =
  {-}2\int_{{-}L/2}^{L/2} \dd x \,\psiinf(\cscale x)
  V(\cscale x) \psiinf(\cscale x - \cscale L) \\
  + \OO (\ee^{\frac32 \ii \cscale p_\infty L}) \,.
\label{eq:DeltaE-1D-int}
\end{multline}
This result is exactly analogous to the ordinary bound-state result in 1D,
except that the coordinate $x$ is replaced with $x\cscale = x \ee^{\ii \phi}$.
Using integration by parts, we can write
\begin{multline}
 \Delta E(L)
 = \frac{\cscale^*}{\mu}\Bigg[
  \psiinf(\cscale  x- \cscale L)\frac{\dd}{\dd x}\psiinf(\cscale x) \\
  \null - \psiinf(\cscale x)\frac{\dd}{\dd x}\psiinf(\cscale x - \cscale L)
 \Bigg]\Bigg|_{{-}L/2}^{L/2}
 + \OO\left(\ee^{\frac32 \ii \cscale p_\infty L}\right) \,.
\end{multline}
Asymptotically, \ie, for $\abs{\cscale x} = \abs{x}$ outside the range $R$ of
the short-range interaction, the complex-scaled infinite-volume resonance wave
function can be written as
\begin{equation}
 \psiinf(\cscale x) = \gamma_\infty \exp( \ii \cscale p_\infty x)
\end{equation}
where $\gamma_\infty$ is the resonance analog of the ANC for bound states.
Inserting this and using that our assumption of even parity implies
$\psiinf (\cscale x-\cscale L) \to \gamma_\infty
\exp( \ii \cscale p_\infty(L-x))$,
we arrive at
\begin{multline}
 \Delta E(L) =
 {-}\frac{2\kappa \gamma_\infty^2}{\mu}
 \exp( \ii \cscale p_\infty L) +\OO(\ee^{\frac32 \ii \cscale p_\infty L}) \\
 ={-} \ee^{ \ii p_\infty  L(\cscale-1)} \left(
  \frac{2\kappa\gamma_\infty^2}{\mu} \exp(\ii p_\infty L)
 \right) + \OO\left(\ee^{\frac32\ii \cscale p_\infty L}\right) \,.
\end{multline}

For a three-dimensional $S$-wave state, we can follow exactly the same
procedure, with only minor technical changes to account for the cubic
boundary condition and the occurrence of partial
derivatives~\cite{Konig:2011nz,Konig:2011ti}.
The result that we obtain for the resonance energy shift from this
procedure is
\begin{equation}
 \Delta E(L) = \frac{3 \gamma_\infty^2}{\mu}
 \frac{\exp(\ii \cscale p_\infty L)}{\cscale L}
 + \OO\left(\ee^{\sqrt{2}\ii \cscale p_\infty L}\right) \,.
\label{eq:LOform}
\end{equation}

We note that Eq.~\eqref{eq:LOform} can be obtained from the bound-state
relation without complex scaling, Eq.~\eqref{eq:LOform-BS}, by rotating
the box size as $L \to \cscale L$ and replacing the binding momentum
$\kappa$ with ${-}\ii  p_\infty$.
We also point out that the complex-scaled form of the volume dependence indeed
still applies to bound states calculated with complex scaling, \ie,
Eq.~\eqref{eq:LOform} remains valid for $p_\infty = \ii\kappa$ with
real $\kappa > 0$ (since bound-state energies remain real under
complex scaling)~\cite{Balslev:1971vb,Moiseyev:1998aa}.

\subsubsection{Subleading corrections}

The imaginary part of the exponent in Eq.~\eqref{eq:LOform}
gives $\Delta E(L)$ an oscillatory behavior as a function of $L$.
While the subleading terms arising from $\braket{\psi_\phi'|\eta}$ are
exponentially suppressed as far as the \emph{magnitude} of $\Delta E(L)$ is
concerned, these contributions can be significant to simultaneously describe
the real and imaginary parts of the energy shift with good accuracy.
We therefore derive in the following the explicit form of the volume
dependence including the first subleading corrections.
Following Ref.~\cite{Guo:2020ikh}, we define a complex-scaled finite-volume
Green's function as
\begin{equation}
 G_{\cscale L}(\zeta \vecr,E)
 = \frac{1}{\cscale  L^3}\sum_{\vecp\in\Gamma_L}
 \frac{\ee^{\ii \vecp \cdot \vecr}}
 {(\zeta^*)^2 \vecp^2 - 2\mu E} \,,
\end{equation}
with
$\Gamma_L = \{\vecp : \vecp = \frac{2\pi \vecn}{L}, \vecn\in \mathbb{Z}^3\}$.
This function satisfies the finite-volume Helmholtz equation
\begin{equation}
 \left[(\zeta^*)^2 \Delta + 2\mu E\right] \cscale  G_{\cscale L}(\zeta \vecr,E)
 = {-}\sum_{\vecn \in \mathbb{Z}^3}\delta(\vecr + \vecn L) \,,
\end{equation}
and it is related to the Lüscher's standard finite-volume Green's function
$G_{L}(\vecr,E)$~\cite{Luscher:1990ux} by the following relation:
\begin{equation}
 G_{\cscale L}(\zeta\vecr,E) = \zeta G_{L}(\vecr,\zeta^2 E) \,.
\end{equation}
The above equality for the Green's functions relates the complex scaling
of the coordinate to a scaling of the energy and it thereby allows us to
apply the analysis of Ref.~\cite{Luscher:1990ux} to the complex-scaled system.
Our starting point is the relation between the scattering ($S$) matrix and
finite-volume energy levels, which for the $A_1^+$ irreducible representation
of the cubic group, truncated to $S$-wave contributions, reads
\begin{equation}
 \ee^{2\ii \delta_0(p)}
 = \frac{\mathcal Z_{00}(1;q^2) + \ii \pi^{3/2} q}
 {\mathcal Z_{00}(1;q^2) - \ii \pi^{3/2}q} \,,
\label{eq:S-matrix-l0}
\end{equation}
with $\mathcal{Z}_{00}$ denoting Lüscher's zeta function~\cite{Luscher:1990ux}.

While we followed Lüscher in writing the infinite-volume $S$ matrix in terms
of a scattering phase shift $\delta_0(p)$ in
Eq.~\eqref{eq:S-matrix-l0}, we note that localized states in the spectrum,
\ie, bound states and resonances that are exponentially decaying after complex
scaling, the analytically continued $S$ matrix has corresponding poles at
complex momenta $p$.
To find the finite volume dependence of these states, we can expand
Eq.~\eqref{eq:S-matrix-l0} around the infinite-volume limit, following
Ref.~\cite{Beane:2003da}.
We start by writing the $S$ matrix in the form
\begin{equation}
 \ee^{2\ii \delta_0(p)}
 = \frac{p\cot\delta_0(p) +\ii p}{p\cot\delta_0(p) - \ii p}
\end{equation}
and consider $K_0(p) \equiv p\cot\delta_0(p)$ as a function of complex $p$.
The quantization condition~\eqref{eq:S-matrix-l0} then takes the simpler
form
\begin{equation}
 K_0(p) = \frac{\sqrt{4\pi}}{\pi L} \mathcal Z_{00}(1;q^2) \,,
\label{eq:pcotan}
\end{equation}
and the condition for a pole in the $S$ matrix becomes $K_0(p) = \ii p$.

We now regard $p = p(L)$ as the volume-dependent momentum corresponding to the
resonance pole, related to the resonance energy $E = E(L)$ via
$p = \sqrt{2\mu E}$.
In infinite volume, the pole is at $p = p_\infty = \sqrt{2\mu E_\infty}$.
As discussed above, we can apply complex scaling now directly to
Eq.~\eqref{eq:pcotan} to derive the desired volume dependence, \ie,
we consider $p \to \cscale p$ (which trivially implies $q \to \cscale q$).
Expanding the left side of Eq.~\eqref{eq:pcotan} around the (complex-scaled)
infinite-volume limit, using $K_0(\cscale p)= K_0(\cscale p(E))$ and evaluating
the expansion at $E = E(L)$, we get
\begin{multline}
 K_0(\cscale p) = K_0(\cscale p_\infty)
 + K_0'(\cscale p_\infty) \frac{\cscale\mu}{p_\infty}
 \left(E(L) - E(\infty)\right) \\
 + \OO\left(\left(E(L) - E(\infty\right))^2\right) \,.
\label{eq:K0-expanded}
\end{multline}
We use the prime here to denote the derivative of $K_0$ with respect to its
(momentum) argument and the factor in Eq.~\eqref{eq:K0-expanded} arises from
$\dd(\cscale p)/\dd E\big|_{p=p_\infty}$.
The purpose of performing the expansion in terms of the energy is that the
finite-volume energy shift $E(L) - E_\infty = \Delta E(L)$ appears explicitly
in Eq.~\eqref{eq:K0-expanded}.
Note also that via the pole condition in infinite volume we have
$K_0(\cscale p_\infty) = \ii \cscale p_\infty$.

The right-hand side of Eq.~\eqref{eq:pcotan} contains Lüscher's zeta function.
We can analytically continue this $\mathcal Z_{00}(1;q^2)$ to the full complex
plane of $q$~\cite{Elizalde:1997jv}, and make use of the following series
expansion:
\begin{multline}
 \frac{\sqrt{4\pi}}{\pi L} \mathcal Z_{00}(1;\cscale^2 q^2)
 = \ii \cscale p
 + \sum_{\vec n\in \mathbb Z^3}^{}{}^{\!\!'}\,
 \frac{\exp(2\pi\ii  \abs{\vec n} \cscale q)}{\abs{\vec n} L}\,,
\label{eq:zeta-expanded}
\end{multline}
where the prime on the sum means that $\vecn =0 $ is to be excluded.
Note that $p=p(L)$ and $q=q(L)$ here.
Combining Eqs.~\eqref{eq:K0-expanded} and~\eqref{eq:zeta-expanded}, and
noting that
\begin{equation}
 p(L) - p_\infty \equiv \Delta p(L) = \frac{\mu}{p_\infty}\Delta E(L) \,,
\end{equation}
or equivalently expanding $p(L) = p(E(L))$ around $L=\infty$ similar to
Eq.~\eqref{eq:K0-expanded}, we obtain
\begin{multline}
 \frac{\cscale\mu}{p_\infty}\left[K_0'(\cscale p_\infty)-\ii\right]
 \Delta E(L) = \sum_{\vec n\in \mathbb Z^3}^{}{}^{\!\!'}\,
 \frac{\exp(2\pi\ii  \abs{\vec n} \cscale q)}{\abs{\vec n} L} \\
 + \OO\left((\Delta E)^2\right) \,,
\end{multline}
and ultimately we have
\begin{multline}
 \Delta E(L)
 = \frac{6 p_\infty}{\cscale\mu \left[K_0'(\cscale p_\infty)-\ii\right] L}
 \times \Bigg[ \exp(\ii \zeta p_\infty L) \\
 \null + \sqrt{2} {\exp(\ii \sqrt{2} \zeta p_\infty L)}
 + \frac{4}{3\sqrt{3}} \exp(\ii \zeta \sqrt{3}p_\infty L) \Bigg] \\
 \null + \OO\left(\ee^{\ii 2 \cscale  p_\infty L}\right) \,.
\label{eq:NLOform}
\end{multline}
We find that this method generates the first subleading terms contributing to
$\Delta E(L)$, as desired, and we note that also yet higher-order subleading
terms can be derived by using this expansion.
The $\OO((\Delta E)^2)$ term in Eq.~\eqref{eq:K0-expanded} then appears
together with $\OO\left(\ee^{\ii 2 \zeta  p_\infty L}\right)$ terms from the
expansion of the zeta function, and at this point the number of unknown
parameters increases.
For more details, we refer to Ref.~\cite{Furnstahl:2013vda}, where the $S$
matrix is expanded in the context of deriving extrapolations for truncated
harmonic oscillator bases.
This basis truncation can be related to an effective spherical hard-wall
boundary and can thus be studied with techniques similar to what we have used
here (see also Refs.~\cite{Furnstahl:2012qg,More:2013rma}).

The prefactor in Eq.~\eqref{eq:NLOform} contains the unknown
quantity $K_0'(\cscale p_\infty)$.
Overall, we can relate the prefactor to the residue of the $S$ matrix at the
resonance pole.
For bound states, we would obtain the (squared) asymptotic
normalization constant (ANC)~\cite{Faldt:1996sd}, and this relation has
been extended to resonances, where the ANC becomes proportional to the
resonance width~\cite{Mukhamedzhanov:2023ecg}.
In light of this correspondence, we can identify
\begin{equation}
 \gamma_\infty^2 = \frac{2p_\infty}{K_0'(\cscale p_\infty)-\ii}
\label{eq:ANC-K}
\end{equation}
and write the final form of the volume dependence as
\begin{multline}
 \Delta E(L)
 = \frac{3\gamma_\infty^2}{\mu\cscale L} \times \Bigg[
 \exp(\ii \zeta p_\infty L)
 + \sqrt{2}{\exp(\ii \sqrt{2} \zeta p_\infty L)} \\
 \null +\frac{4}{3\sqrt{3}} \exp(\ii \zeta \sqrt{3}p_\infty L) \Bigg]
 + \OO\left(\ee^{\ii 2 \cscale p_\infty L}\right) \,,
\label{eq:NLOform-final}
\end{multline}
establishing also the connection with the leading form~\eqref{eq:LOform}.

This derivation can also be generalized to higher angular momenta.
In particular, $P$-wave (angular momentum $l=1$) bound states in infinite
volume typically fall into $T^-_1$ cubic representation.
We assume here that this remains true for resonances because like bound states
these correspond to isolated $S$-matrix poles.
According to Ref.~\cite{Luscher:1990ux}, the following quantization condition
holds in this channel:
\begin{equation}
 K_1(p)\equiv p \cot \delta_1(p)
 = \frac{\sqrt{4\pi}}{\pi L} \mathcal{Z}_{00}(1;q^2) \,.
\label{eq:pcotan-l1}
\end{equation}
Note that this relation is still using $\mathcal{Z}_{00}$, with higher-order
zeta functions contributing to $T^-_1$ only once $l \geq 3$ waves are
considered.
For localized states with angular momentum $l$, the residues of the
corresponding $S$-matrix poles come with a factor
$(-1)^l$~\cite{Faldt:1996sd,Mukhamedzhanov:2023ecg}.
We therefore write the $P$-wave analog of Eq.~\eqref{eq:ANC-K} as
\begin{equation}
 \gamma_\infty^2 = {-} \frac{2p_\infty}{K_1'(\cscale p_\infty)-\ii} \,,
\end{equation}
and using that, we arrive at
\begin{multline}
 \Delta E(L)
 = {-}\frac{3\gamma_\infty^2}{\mu\cscale L} \times \Bigg[
  \exp(\ii \zeta p_\infty L)
  + {\sqrt{2}}{\exp(\ii \sqrt{2} \zeta p_\infty L)} \\
  \null + \frac{4}{3\sqrt{3}} \exp(\ii \zeta \sqrt{3}p_\infty L) \Bigg]
 + \OO\left(\ee^{\ii \cscale 2 \cscale p_\infty L}\right) \,.
\label{eq:NLOform-final-P}
\end{multline}
In particular, for $p_\infty = \ii\kappa$ and without complex scaling ($\phi\to0
\implies \cscale\to1$), the leading term in Eq.~\eqref{eq:NLOform-final-P}
recovers the known $P$-wave result for bound
states~\cite{Konig:2011nz,Konig:2011ti}.

\section{Numerical implementation}
\label{sec:Numerical}

In order to numerically test the relations derived in the previous section, we
use the \emph{finite-volume discrete variable representation (FV-DVR)} as
described in Refs.~\cite{Klos:2018sen,Dietz:2021haj,Konig:2022cya}.
We refer to those reference for details about the method and its efficient
numerical implementation and focus here only on the adaption the basic building
blocks to support uniform complex scaling within the FV-DVR.

The starting point for the FV-DVR is a plane-wave basis
\begin{equation}
 \phi_j^{(L)}(x)
 = \frac{1}{\sqrt{L}} \exp\left(\ii\frac{2\pi j}{L} x\right) \,,
\label{eq:PW-basis}
\end{equation}
where $L$ as before is the size of the periodic volume and the index $j$ runs
from ${-}N/2$ to $N/2$ for even number of modes $N>2$.
The $x$ in Eq.~\eqref{eq:PW-basis} denotes the relative coordinate describing
a two-body ($n=2$) system in one dimension ($d=1$).
As in the derivation of the resonance volume dependence in the previous
section, it is convenient to initially discuss this simple scenario.
For a set of equidistant points $x_k \in [{-}L/2,L/2)$ with associated weights
$w_k = L/n$ (defining together a simple trapezoidal integration rule), DVR
states are constructed from the $\phi_j^{(L)}(x)$ by means of a unitary
transformation~\cite{Groenenboom:2001aa}
\begin{equation}
 \psi_k(x) = \dvrsum{j}{N} \mathcal{U}^*_{kj} \phi_j(x) \,,
 \label{eq:psi-dvr}
\end{equation}
with $\mathcal{U}_{ki} = \sqrt{w_k} \phi_i(x_k)$.
The index $k$ in Eq.~\eqref{eq:psi-dvr} covers the same range of integers as
the $j$ labeling the original plane-wave modes, and $\psi_k(x)$ is a
wave function peaked at $x_k$.
In order to apply the method, a generic Hamiltonian as in
Eq.~\eqref{eq:H-generic} is expanded within the basis spanned by the DVR
states $\ket{\psi_k} \equiv \ket{k}$.
The kinetic-energy operator for the one-dimensional two body system,
expressed in coordinate space, is, up to a prefactor ${-}1/(2\mu)$ simply
a second derivative with respect to $x$, and more generally it takes
the form as given in Eq.~\eqref{eq:K},
featuring combinations of partial derivatives with respect to the coordinates.
For each such individual derivative, DVR matrix elements can be written down
explicitly in closed form,
\begin{equation}
 \bra{k} \partial \ket{l} = \frac{\pi (-1)^{k-l}}{L}
  \begin{cases}
    {-}\ii & \text{if } k=l\\
    \frac{\exp \left[ {-}\ii\frac{\pi (k-l)}{N} \right]}
    {\sin{\frac{\pi (k-l)}{N}}}
    & \text{otherwise}
  \end{cases} \,,
\label{eq:q-braket}
\end{equation}
and from this one directly obtains an explicit representation for $H_0$.
For local potentials, the DVR has the convenient property that these
are represented by diagonal matrices,
\begin{equation}
 \bra{k} V \ket{l} \approx V(x_k) \delta_{kl} \,,
\label{eq:V-diagonal}
\end{equation}
with a very good approximate identity that becomes exact in the limit
$N\to\infty$.

For arbitrary number of particles $n$ and spatial dimensions $d$, DVR states
can be written as
\begin{equation}
 \ket{s}
 = \ket{(k_{{1,1}},\ldots k_{{1,d}}),\ldots ,(k_{{n-1,1}},\ldots k_{{n-1,d}})} \,,
\label{eq:s}
\end{equation}
and the corresponding wave functions are simply tensor products of 1D modes:
\begin{equation}
 \psi_s(\underline{x})
 = \braket{\underline{x}|s}
 = \prod_{
  \substack{i=1,\cdot\cdot n-1 \\c=1,\cdot\cdot d}
 } \psi_{k_{i,c}}(x_{i,c}) \,.
\end{equation}
We note that the $\ket{s}$ can in addition include discrete quantum numbers
such as spin and isospin but neglect these here for simplicity.
The $d$-dimensional kinetic-energy operator for $n$ particles can then be
constructed as
\begin{equation}
 H_0 = K \oplus K \oplus \ldots \oplus K \quad \text{($d$ times)} \,,
\label{eq:H_0}
\end{equation}
where $\oplus$ denotes the Kronecker sum~\cite{KroneckerSumWolfram}, and
$K$ is the 1D kinetic energy operator given by restricting the sum
over $c$ in Eq.~\eqref{eq:K} to just one term.
For example, for a two-body system in $d=3$ dimensions this definition
amounts to a sparse DVR matrix with entries
\begin{spliteq}
 &\bra{k_{1,1}, k_{1,2}, k_{1,3}} H_0 \ket{l_{1,1}, l_{1,2}, l_{1,3}} \\
 &\hspace{2em}= \bra{k_{1,1}} K \ket{l_{1,1}}
 \delta_{k_{1,2},l_{1,2}} \delta_{k_{1,3},l_{1,3}} \\
 &\hspace{4em}+ \bra{k_{1,2}} K \ket{l_{1,2}}
 \delta_{k_{1,1},l_{1,1}} \delta_{k_{1,3},l_{1,3}} \\
 &\hspace{4em}+ \bra{k_{1,3}} K \ket{l_{1,3}}
 \delta_{k_{1,1},l_{1,1}} \delta_{k_{1,2},l_{1,2}} \,,
\end{spliteq}
\ie, it can be constructed in terms of the 1D matrix elements, and this remains
true for $n>2$.
Similarly, the evaluation of local two-body interactions generalizes
straightforwardly to $d>1$ and $n>2$.
For more than two particles, there is a pairwise two-body interaction
for each pair, as discussed above Eq.~\eqref{eq:V-xixj}.
In the DVR, for each such pairwise interaction there are appropriate
Kronecker deltas for the spectator
particles~\cite{Klos:2018sen,Konig:2020lzo}.

As per our previous discussion in Sec.~\ref{sec:ComplexScaling}, complex scaling
in simple relative coordinates, and therefore for the DVR, is applied
simultaneously to each coordinate and component.
For the kinetic-energy term in the DVR basis we therefore only need to
adjust the 1D two-body matrix elements to implement complex scaling, and
everything else then follows from that.
Specifically, a factor $\cc{\cscale}$ in included in Eq.~\eqref{eq:q-braket},
leading to the previously derived scaling of $H_0$ with a factor
$(\cc{\cscale})^2$.
Similarly, for local two-body interactions we simply apply the scaling
to each relative separation when evaluating the potential matrix elements,
and Eq.~\eqref{eq:V-diagonal} (and its generalization to $d$ dimensions and
$n$ particles) implies that this carries over directly to the DVR.

\section{Examples}
\label{sec:Examples}

We use the complex-scaled FV-DVR discussed in the previous section to
study several explicit examples.
Our goal is to obtain infinite-volume energies
$E_\infty = {p_\infty^2}/{2\mu}$ from a set of calculations at finite $L$.
To that end, we can fit the numerical data to the functional forms derived in
Sec.~\ref{sec:VolumeDependence}, thereby determining the unknown variables
$p_\infty$ and $\gamma_\infty$ in Eqs.~\eqref{eq:LOform}
and~\eqref{eq:NLOform-final} (or the corresponding $P$-wave forms).
In order to use standard least-squares minimization that is typically applied
to real functions of real parameters, we separate $\Rp\, E(L)$ and $\Ip\, E(L)$ and
fit then both of them simultaneously, while also expanding the complex
parameters $\{p_\infty,\gamma_\infty\}$ into
$\{\Rp\, p_\infty,\Ip\, p_\infty,\Rp\, \gamma_\infty,\Ip\, \gamma_\infty\}$.

\subsection{$S$-wave resonance}

\begin{figure}[htbp]
\centering
\includegraphics[width=1\linewidth]{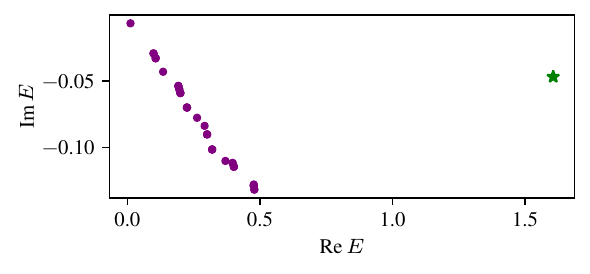}
\caption{%
Complex-$E$ spectrum at $L=20$ for the two-body potential given by
Eq.~\eqref{eq:potentialA} showing 40 eigenvalues with the largest imaginary
parts.
The resonance of interest is highlighted with a star symbol.
\label{fig:2b-S-res-snapshot}
}
\end{figure}

From Ref.~\cite{Klos:2018sen} it is known that the potential,
\begin{equation}
 V(r) = 2\exp[{-}\left(\frac{r-3}{1.5}\right)^2]
\label{eq:potentialA}
\end{equation}
generates an $S$-wave resonance at $E_\infty = 1.606(1) - \ii 0.047(2)$
for a two-body system with $m=2\mu=1$, using natural units $\hbar=c=1$.
We use this potential here in an FV-DVR calculation with a DVR basis size
$N=96$ and a complex-scaling angle of $\phi=\pi/24$.
For this calculation, we determine the finite-volume energy spectrum by
selecting states with largest imaginary part.
As a representative example for what this (partial) spectrum looks like,
we show in Fig.~\ref{fig:2b-S-res-snapshot} the loci of the 40 energy
levels (counting degeneracies) with largest imaginary part in an $L=20$ box.
Since in this case we know the exact infinite-volume energy $E$ for the
resonance of interest, we can easily select from from the spectrum the value
that is closest to it.
In practical applications, where the expected result is not known in advance,
one can repeat the calculation for several rotation angles $\phi$
and identify as physical resonances the levels that do not move significantly
under this angle variation, as predicted by the Balslev-Combes
theorem~\cite{Moiseyev:1978aa,Balslev:1971vb}.

%%%%%%%%%%%%%%%%%%%%%%%%%%%%%%%%%%%%%%%%%%%%%%%%%%%%%%%%%%%%%%%%%%%%%%%%%%%%%%%%
%%%%%%%%%%%%%%%%%%%%%%%%%%%%%%%%%%%%%%%%%%%%%%%%%%%%%%%%%%%%%%%%%%%%%%%%%%%%%%%%

\begin{figure}[htbp]
\centering
\includegraphics[width=1\linewidth]{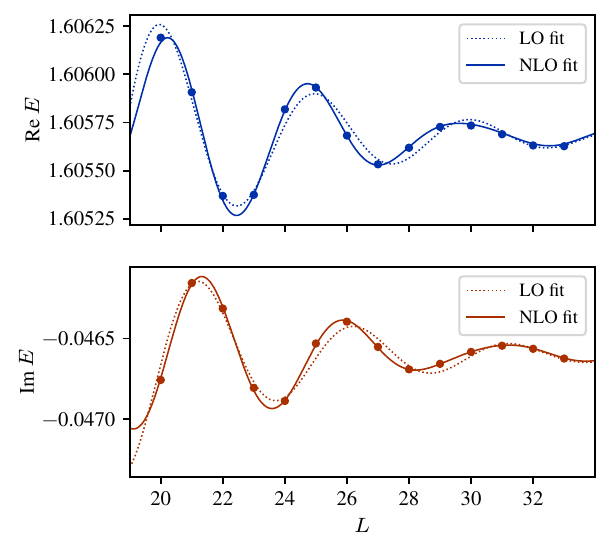}
\caption{%
Finite-volume spectrum for the $S$-wave resonance for the potential given by
Eq.~\eqref{eq:potentialA}.
The real (imaginary) part of the energy is shown in the upper (lower) panel.
For each, we show the result of fitting the volume dependence at LO (dotted
line) and NLO (solid line).
\label{fig:2b-S-res}
}
\end{figure}

Being able to identify the resonance state of interest, we can repeat the
calculation for a range of volumes and perform the fits as described at
the beginning of this section.
The result is shown in Fig.~\ref{fig:2b-S-res}.
For comparison, we fit both the leading-order (LO) form of the volume
dependence, Eq.~\eqref{eq:LOform}, as well as the ``NLO'' form given in
Eq.~\eqref{eq:NLOform-final}.
From the LO fit we obtain $E_\infty = 1.605676(13) - \ii 0.046603(13)$, while
the NLO fit gives $E_\infty = 1.6056798(27) - \ii 0.0465947(27)$, in good
agreement with the known value for this resonance.
The uncertainties quoted here for our calculation are the standard errors
reported by the fitting routine.

%%%%%%%%%%%%%%%%%%%%%%%%%%%%%%%%%%%%%%%%%%%%%%%%%%%%%%%%%%%%%%%%%%%%%%%%%%%%%%%%
%%%%%%%%%%%%%%%%%%%%%%%%%%%%%%%%%%%%%%%%%%%%%%%%%%%%%%%%%%%%%%%%%%%%%%%%%%%%%%%%

Moreover, instead of varying the volume in order to extrapolate to $L=\infty$,
it is also possible to keep $L$ fixed and then fit the energy as a function of
the complex-scaling rotation angle $\phi$, restricted by the condition that
$\phi > {-}\arg E/2$.

\begin{figure}[htbp]
  \centering
  \includegraphics[width=1\linewidth]{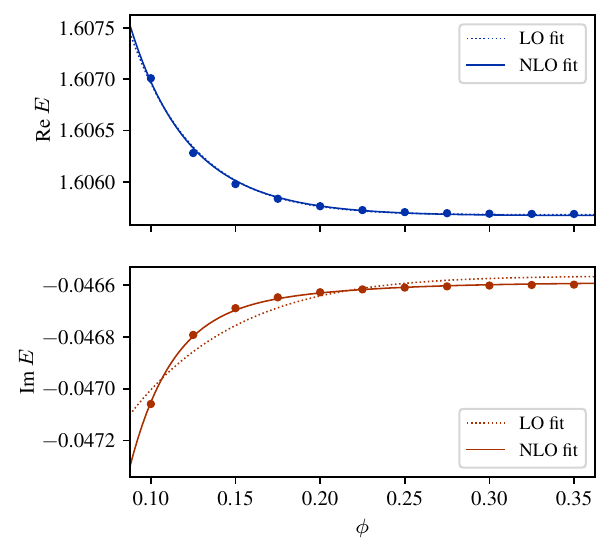}
  \caption{%
  Finite-volume energy, as a function of the complex-scaling
  angle $\phi$, for the $S$-wave resonance for the potential given by
  Eq.~\eqref{eq:potentialA}.
  The real (imaginary) part of the energy is shown in the upper (lower) panel.
  For each, we show the result of fitting the volume dependence at LO (dotted
  line) and NLO (solid line).
  \label{fig:2b-S-res-phi}
  }
\end{figure}

To demonstrate this, we perform another FV-DVR calculation for the same system,
with a constant box size of $L=20$, a DVR basis size of $N=80$, but
varying the angle $\phi$ in the range shown in Fig.~\ref{fig:2b-S-res-phi}.
Curve fitting is performed as described previously, except that now the
independent variable is $\phi$ instead of $L$.
We obtain $E_\infty = 1.605681(13) - \ii 0.046565(13)$ from the LO fit and
$E_\infty = 1.605673(6) - \ii 0.046591(6)$ from the NLO fit.
Noting that the standard errors we report from the fitting routine are only
part of the actual theoretical uncertainty, which in particular also arises from
unknown higher-order terms in the analytical form of the angle/volume
dependence, this value is in reasonable agreement with the result from varying
$L$ and with the reference value.

%%%%%%%%%%%%%%%%%%%%%%%%%%%%%%%%%%%%%%%%%%%%%%%%%%%%%%%%%%%%%%%%%%%%%%%%%%%%%%%%
%%%%%%%%%%%%%%%%%%%%%%%%%%%%%%%%%%%%%%%%%%%%%%%%%%%%%%%%%%%%%%%%%%%%%%%%%%%%%%%%

\subsection{$P$-wave resonance}

\begin{figure}[htbp]
\centering
\includegraphics[width=1\linewidth]{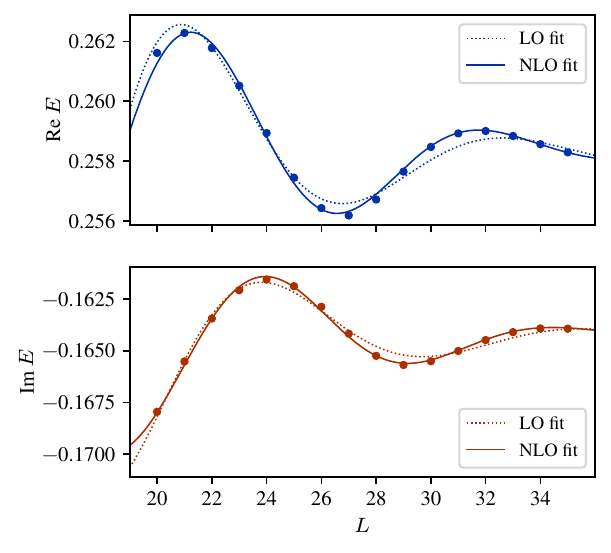}
\caption{%
Finite-volume spectrum for the $P$-wave resonance for the potential given by
Eq.~\eqref{eq:potentialB}.
The real (imaginary) part of the energy is shown in the upper (lower) panel.
For each, we show the result of fitting the volume dependence at LO (dotted
line) and NLO (solid line).
\label{fig:2b-P-res}
}
\end{figure}

To study a $P$-wave example, we use the potential
\begin{equation}
 V(r) = {-}10 \exp({-}r^2) \,,
 \label{eq:potentialB}
\end{equation}
which we find to support a resonance at $E_\infty = 0.25822632 - \ii 0.16432586$
from a momentum-space calculation with complex scaling (see
Ref.~\cite{Yapa:2023xyf} for details) with the momentum cutoff and the mesh
resolution increased until the value converged to the quoted precision.
The volume dependence for this state, calculated with with a DVR basis size of
$N=96$ and a complex-scaling angle of $\phi=\pi/6$, is shown
Fig.~\ref{fig:2b-P-res}.
The LO fit for this resonance yields
$E_\infty = 0.25817(7) - \ii 0.16431(7)$, while at NLO we obtain
$E_\infty = 0.258257(31) - \ii 0.164315(31)$.
Both results agree well with the reference value.
We see a marginal improvement in this case at NLO, which is more noticeable in
Fig.~\ref{fig:2b-P-res}: clearly the fit residuals are reduced when using the
NLO volume dependence (solid line in the figure).

%%%%%%%%%%%%%%%%%%%%%%%%%%%%%%%%%%%%%%%%%%%%%%%%%%%%%%%%%%%%%%%%%%%%%%%%%%%%%%%%
%%%%%%%%%%%%%%%%%%%%%%%%%%%%%%%%%%%%%%%%%%%%%%%%%%%%%%%%%%%%%%%%%%%%%%%%%%%%%%%%

\begin{figure}[htbp]
  \centering
  \includegraphics[width=1\linewidth]{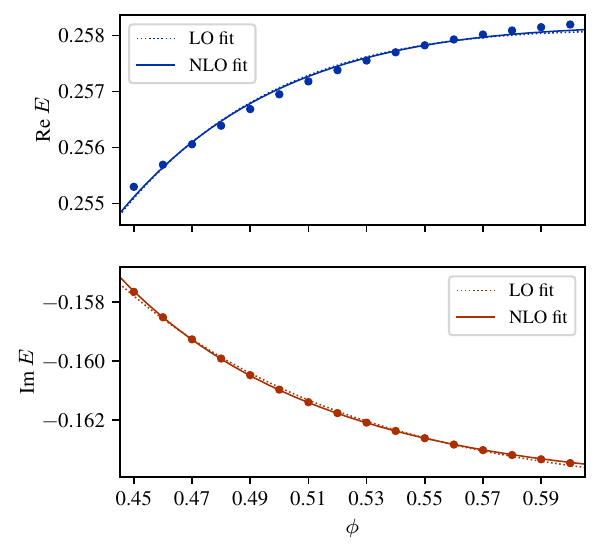}
  \caption{%
  Finite-volume energy, as a function of the complex-scaling
  angle $\phi$, for the $P$-wave resonance for the potential given by
  Eq.~\eqref{eq:potentialB}.
  The real (imaginary) part of the energy is shown in the upper (lower) panel.
  For each, we show the result of fitting the volume dependence at LO (dotted
  line) and NLO (solid line).
  \label{fig:2b-P-res-phi}
  }
\end{figure}

As we did for the $S$-wave resonance, we repeat the calculation in a fixed box
with $L=25$, using a DVR basis size of $N=96$, and vary the complex-scaling
angle $\phi$, as shown in Fig.~\ref{fig:2b-P-res-phi}.
For this particular resonance, which is very wide (the imaginary part of the
energy has a magnitude that is more than 60\% the magnitude of the real
part), obtaining very accurate results from fitting the $\phi$ dependence is
challenging.
Due to the large width, which leads to a large $\arg E$, the minimum complex
scaling angle is of the order $0.4$ radians, leaving only a relatively narrow
window to vary $\phi$ in.
Moreover, towards the smaller end of the permissible window, complex scaling
only induces a rather weakly decaying behavior of the wave function, and
therefore higher exponential terms $\OO(\ee^{\ii 2 \cscale  p_\infty L})$ are
not particularly strongly suppressed.
The effect of this can be seen most noticeably in the upper panel of
Fig.~\ref{fig:2b-P-res-phi}, where we show the fit result for the real part of
the energy.
The curve fitting was otherwise performed as before, to obtain
$E_\infty = 0.25782(4) - \ii 0.16447(4)$ as the LO result and
$E_\infty = 0.258017(28) - \ii 0.164225(28)$ as the NLO result.
In this case it is particularly obvious that the standard fit errors alone
underestimate the true uncertainty, but we point out that nevertheless the NLO
result agrees with the $L$-based fit to better than $0.5\%$ for the real part
and to better than $0.1\%$ for the imaginary part.

%%%%%%%%%%%%%%%%%%%%%%%%%%%%%%%%%%%%%%%%%%%%%%%%%%%%%%%%%%%%%%%%%%%%%%%%%%%%%%%%
%%%%%%%%%%%%%%%%%%%%%%%%%%%%%%%%%%%%%%%%%%%%%%%%%%%%%%%%%%%%%%%%%%%%%%%%%%%%%%%%

\subsection{$S$-wave bound state}

As mentioned in Sec.~\ref{sec:VolumeDependence}, the volume dependence derived
in this work is valid not only for resonances, but also for bound states
calculated with complex scaling.
In infinite volume, bound-state energies remain real under complex scaling, but
for $L<\infty$ our analytical calculation predicts that they in general acquire
a non-zero imaginary part.
The utility in this formalism for bound states lies in the fact that
extrapolation can be performed for a constant $L$ while varying $\phi$, which is
what we opt to do here, noting that fitting the $L$ dependence (without complex
scaling) is known to work well for bound states (see for example
Refs.~\cite{Konig:2011nz,Konig:2011ti,Konig:2017krd,Yu:2022nzm}).
In this case, unlike resonances, the available range of angles is no longer
restricted by the condition that $\phi > {-}\arg E/2$.

\begin{figure}[htbp]
\centering
\includegraphics[width=1\linewidth]{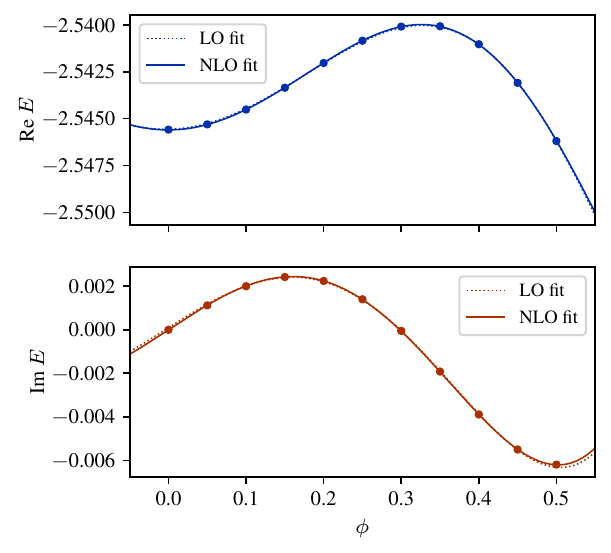}
\caption{%
Finite-volume energy, as a function of the complex-scaling
angle $\phi$, for the $S$-wave bound state generated by the potential
given by Eq.~\eqref{eq:potentialB}.
The real (imaginary) part of the energy is shown in the upper (lower) panel.
For each, we show the result of fitting the volume dependence at LO (dotted
line) and NLO (solid line).
\label{fig:2b-S-bound-phi}
}
\end{figure}

As a concrete example, we look at the $S$-wave bound state with
$E_\infty={-}2.5434016$ (reference value obtained from a momentum-space
calculation, with uncertainty smaller than the given number of digits)
generated by the same potential~\eqref{eq:potentialB} we used to generate
a $P$-wave resonance.
We carry out this calculation using an FV-DVR calculation with a box size of
$L=6$ and a DVR basis size of $N=30$.
The rotation angle is varied in a range as shown in Fig.~\ref{fig:2b-S-bound-phi},
and curve fitting is performed as described previously.
From the LO fit we obtain $E_\infty = {-}2.543428(15) + \ii 0.000022(15)$ and
the NLO yields $E_\infty = {-}2.543406(4) - \ii 0.000001(4)$.
While already at LO the real part is in excellent agreement with the reference
value, we see a marginal improvement at NLO.
The imaginary parts are consistent with zero within the uncertainties reported
by the fitting routine.

We note again that the alternative approach of keeping $\phi$ constant while
varying $L$ is just as valid for bound states. The best choice
for such a calculation would be $\phi=0$. However, setting $\phi=0$ in our
analytical expressions would recover the well-established bound-state L\"uscher
formalism~\cite{Luscher:1985dn,Konig:2011ti}, and therefore,
is not studied in this work.

%%%%%%%%%%%%%%%%%%%%%%%%%%%%%%%%%%%%%%%%%%%%%%%%%%%%%%%%%%%%%%%%%%%%%%%%%%%%%%%%
%%%%%%%%%%%%%%%%%%%%%%%%%%%%%%%%%%%%%%%%%%%%%%%%%%%%%%%%%%%%%%%%%%%%%%%%%%%%%%%%

\subsection{Three-boson resonance}

\begin{figure}[htbp]
\centering
\includegraphics[width=1\linewidth]{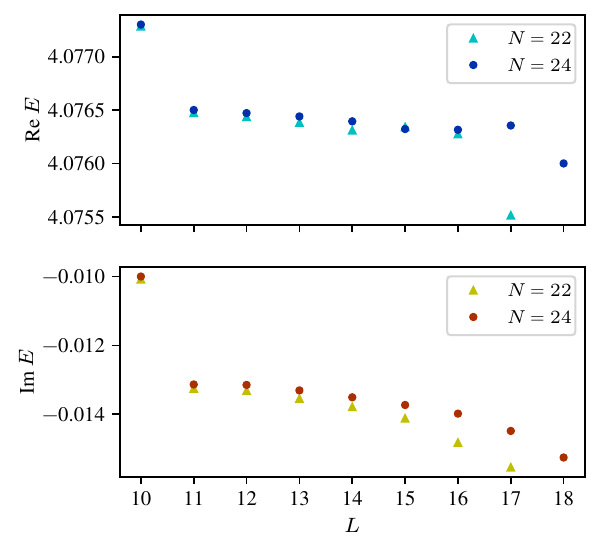}
\caption{%
Finite-volume spectrum for the three-boson resonance generated by the potential
in Eq.~\eqref{eq:potentialA}.
The real (imaginary) part of the energy is shown in the upper (lower) panel.
Since the volume dependence of this state is unknown, this figure does
not include any fitted curves.
\label{fig:3b-res}
}
\end{figure}

Finally, to show that FV-DVR prescription with complex scaling works just
as well beyond the two-body sector, we calculate the finite-volume
three-body spectrum for a system of bosons where the pairwise interaction
between particles is given by the potential~\eqref{eq:potentialA}.
Using the method of avoided crossings, Ref.~\cite{Klos:2018sen} estimates
for this scenario a resonance at $\Rp(E)=4.18(8)$, with an unknown
width.
We study this system with a complex-scaling angle $\phi=\pi/9$, employing
symmetrization to restrict the calculation to bosonic states with positive
parity.
We find indeed a resonance close to the expected position, identified in
the same manner as discussed for two-body resonances.
The volume dependence of this state is shown in Fig.~\ref{fig:3b-res},
where in order to study numerical convergence with the DVR basis size we
compare results for $N=22$ and $N=24$.
From this comparison we conclude that the real part of the energy is well
converged up to at least $L=16$, whereas the imaginary part shows somewhat
larger remaining artifacts due to a lack of (ultraviolet) convergence.

%%%%%%%%%%%%%%%%%%%%%%%%%%%%%%%%%%%%%%%%%%%%%%%%%%%%%%%%%%%%%%%%%%%%%%%%%%%%%%%%
%%%%%%%%%%%%%%%%%%%%%%%%%%%%%%%%%%%%%%%%%%%%%%%%%%%%%%%%%%%%%%%%%%%%%%%%%%%%%%%%

Since we do not know the functional form for the volume dependence of
this three-body state, we cannot use the fitting technique to directly
infer the infinite-volume energy for this resonances.
However, one should expect that similar to two-body resonances the \emph{norm}
of the energy to converges exponentially with increasing $L$, as pointed out
previously in Ref.~\cite{Guo:2020ikh}.
Indeed, noting the inflated vertical axis scale in Fig.~\ref{fig:3b-res},
we point out that compared to the overall magnitude, both the real and
the imaginary part of the energy show only relatively small variations
over the range $L = 11,\ldots ,16$ shown in the figure.
As a very rough estimate for the infinite-volume properties, we merely
take the average of the $N=24$ results over this range to obtain
$E_\infty \approx 4.07641(8) - \ii 0.01347(34)$.
The real part we find is close to the value $\Rp(E)=4.18(8)$ reported in
Ref.~\cite{Klos:2018sen}, although the respective uncertainties do not
quite overlap.
Since we see very little variation of $\Rp(E)$ with $L$ or $N$, we presume
that Ref.~\cite{Klos:2018sen} likely underestimated the uncertainty
stemming from the method of avoided level crossings.

\section{Summary and outlook}
\label{sec:Conclusion}

In this work, we have studied complex scaling in finite periodic boxes
as a framework for studying few-body quantum systems, in particular
systems that host resonances.
We derived explicitly the volume dependence of two-body resonances and bound
states (\ie, energy levels that correspond to isolated $S$-matrix poles in
infinite volume), including the first corrections to the leading behavior.
We furthermore developed a a concrete numerical implementation of
the technique and used this to test the expressions we derived for the
volume dependence with several explicit examples.

Our approach combines two established approaches.
While few-body resonances have been studied in finite-volume without complex
scaling by looking for avoided crossings in the finite-volume
spectrum~\cite{Klos:2018sen,Dietz:2021haj}, that approach can quickly become
numerically expensive, and it does not readily provide access to resonance
widths in general (we note, however, that the ``stabilization
method''~\cite{Hazi:1970aa,Mandelshtam:1993zz,Mandelshtam:1994zz} and
generalizations~\cite{Mueller:1994aa,Kruppa:1999zz,Suzuki:2005wv} can be used to
determine resonance widths indirectly by determining the density of states at a
given box size and fitting it with a Breit-Wigner shape).
Although finite-volume eigenvector continuation~\cite{Yapa:2022nnv} has
been shown to significantly reduce the numerical cost of such studies,
the approach we presented here has the appeal that via complex scaling
resonances can be found in much smaller boxes, and the analytical
expressions we derived then make it possible to directly infer
infinite-volume resonance properties, including the decay width.
Reference \cite{Guo:2020ikh} employs an analytic continuation to imaginary
box sizes in order to study resonances in finite-volume, also including
widths, but that method is still indirect in the sense that it extracts
resonance properties from peaks in transition amplitudes instead of
directly identifying complex energy eigenstates of the finite-volume
Hamiltonian, as we do in this work.

While we have rigorously derived the volume dependence here only
for two-body systems, we expect our results to directly generalize to
few-body states if the dominant decay (or breakup, in the case of bound
states) mode is into two clusters, following the derivation for
bound states without complex scaling~\cite{Konig:2017krd}.
For systems where this is not the case, such as the three-boson example
that we considered in this work, it is still possible to obtain good
approximations to the infinite-volume resonance properties by calculating
in relatively large boxes.

Our findings have applications in various areas of physics, ranging from
cold atoms to nuclear physics.
In particular, it would be interesting to study Efimov trimers (and
associated tetramers)~\cite{Braaten:2004rn,Naidon:2016dpf} in finite volume
with complex scaling, and we also plan to investigate few-neutron
systems using complex scaling in finite volume.
Naturally, our results enable finite-volume studies of resonances in a
a variety atomic nuclei, and developing an extension to systems of charged
particles, as recently done for bound states~\cite{Yu:2022nzm}, will
further broaden the range of systems that our method can be applied to.
Finally, it will be interesting to explore the resonance eigenvector
continuation method developed in Ref.~\cite{Yapa:2022nnv} to study
extrapolations from bound states to resonances in finite volume.

\begin{acknowledgments}
We thank Andrew Andis for useful discussions.
This work was supported in part by the U.S.\ National Science Foundation (Grant
No. PHY--2044632) and by the U.S.\ Department of Energy (DE-SC0024520 --
STREAMLINE Collaboration and DE-SC0024622).
This material is based upon work supported by the U.S.\ Department of Energy,
Office of Science, Office of Nuclear Physics, under the FRIB Theory Alliance,
Award No.\ DE-SC0013617.
Computational resources for parts of this work were provided by the Jülich
Supercomputing Center as well as by the high-performance computing
cluster operated by North Carolina State University.
\end{acknowledgments}

\bibliographystyle{apsrev4-1}

\end{document}